\documentclass[letterpaper]{article} 
\usepackage{aaai24}  
\usepackage{times}  
\usepackage{helvet}  
\usepackage{courier}  
\usepackage[hyphens]{url}  
\usepackage{graphicx} 
\usepackage{tikz}
\usetikzlibrary{positioning}
\usepackage{graphicx}
\usepackage{color}
\usepackage{algorithm}
\usepackage{caption}
\usepackage{subcaption}
\usepackage{multicol}
\usepackage{comment}
\usepackage{tikz}
\usetikzlibrary{positioning}
\usepackage{graphicx}
\usepackage{xspace}
\usepackage{bm}
\usepackage{adjustbox}
\usepackage{tabularx}
\usepackage{color}
\usepackage{algorithm}
\usepackage{listings}
\usepackage{nccmath}
\usepackage{natbib}  
\usepackage{caption} 
\usepackage{algorithm}
\usepackage{algorithmic}

\makeatletter
\DeclareRobustCommand\onedot{\futurelet\@let@token\@onedot}
\def\@onedot{\ifx\@let@token.\else.\null\fi\xspace}

\def\eg{\emph{e.g}\onedot} 
\def\ie{\emph{i.e}\onedot} 
 
\def\etc{\emph{etc}\onedot} 
 
\def\etal{\emph{et al}\onedot}
\makeatother
\usepackage{newfloat}
\usepackage{listings}
\DeclareCaptionStyle{ruled}{labelfont=normalfont,labelsep=colon,strut=off} 
\floatstyle{ruled}
\newfloat{listing}{tb}{lst}{}
\floatname{listing}{Listing}
 \nocopyright 

\usepackage{bibentry}

\title{Topical: Automatic  Repository Tagging using Attention on Hybrid Code Embeddings}
\author {
    Agathe Lherondelle\textsuperscript{\rm 1},
    Varun Babbar\textsuperscript{\rm 1},
    Yash Satsangi\textsuperscript{\rm 1},
    Fran Silavong\textsuperscript{\rm 1},
    Shaltiel Eloul\textsuperscript{\rm 1},\\
    Sean Moran\textsuperscript{\rm 2}
}
\affiliations {
    JPMorgan Chase, 
London, UK\\ 
\textsuperscript{\rm 1}firstname.lastname@jpmchase.com, 
\textsuperscript{\rm 2}sean.j.moran@jpmchase.com
}

\usepackage{bibentry}

\begin{document}

\maketitle

\begin{abstract}
This paper presents Topical, a novel deep neural network for repository level embeddings. Existing methods, reliant on natural language documentation or na\"ive aggregation techniques, are outperformed by Topical's utilization of an attention mechanism. This mechanism generates repository-level representations from source code, full dependency graphs, and script level textual data. Trained on publicly accessible GitHub repositories, Topical surpasses multiple baselines in tasks such as repository auto-tagging, highlighting the attention mechanism's efficacy over traditional aggregation methods. Topical also demonstrates scalability and efficiency, making it a valuable contribution to repository-level representation computation. For further research, the accompanying tools, code, and training dataset are provided at: \url{https://github.com/jpmorganchase/topical}.
\end{abstract}

\section{Introduction}
\label{sec:introduction}
Code hosting platforms like GitHub have transformed the landscape of software development. They offer a collaborative space where developers can explore pertinent code, stay updated with technological trends, and conceptualize new application ideas. For instance, GitHub alone boasts over 200 million repositories\footnote{https://en.wikipedia.org/wiki/GitHub}. Given the sheer volume of open-source repositories, it's paramount to have automated tools that can efficiently search and categorize them. Features such as auto-tagging repositories~\cite{izadi2021topic} with semantic keywords and modeling their topics~\cite{thomas2014studying} are instrumental in handling this vast amount of data.

The field of Machine Learning on Sourcecode (MLOnCode)~\cite{Allamanis18} steps up to the challenge of processing and leveraging source code for practical applications. This discipline has led to numerous innovations, including detecting code duplications~\cite{spinellis2020dataset}, identifying software design patterns~\cite{washizaki2019studying}, improving code quality and auto-generation~\cite{le2020deep}, and even extracting a software developer's skill set directly from their written code~\cite{Allamanis18}. Central to the evolution of these tools is the quest for an optimal representation of code, like code embeddings or repository embeddings~\cite{rokon2021repo2vec,theeten2019import2vec}.

The primary emphasis of Machine Learning on Sourcecode (MLOnCode) research has centered on predictive tasks at either method-level or snippet-level~\cite{feng2020codebert,guo2021graphcodebert,puri21codenet}. Deep neural networks~\cite{goodfellow2016deep} have lately demonstrated remarkable capabilities in comprehending and manipulating method-level code snippets~\cite{Allamanis18, feng2020codebert,guo2021graphcodebert}. This advancement in code embedding can be associated with the integration of transformers equipped with attention mechanisms in natural language models, such as BERT~\cite{devlin-etal-2019-bert} and its related iterations~\cite{peters-etal-2018-deep,Radford2018ImprovingLU,raffel2020exploring}. Comparable neural models have also been adapted for the purpose of source code embedding~\cite{kanade2020learning,feng2020codebert,Svyatkovskiy20,Buratti20,guo2021graphcodebert,karampatsis2020scelmo,husain2020codesearchnet}. These models have achieved striking results at either the script or method level. However, only a minority of these works~\cite{izadi2021topic,rokon2021repo2vec} confront the challenge of aggregating information from numerous scripts to formulate a repository-level representation. The methods that tackle the task of computing repository-level representations~\cite{zhang2019higitclass,izadi2021topic,rokon2021repo2vec} primarily depend on the natural language documentation of the code, such as READMEs and other accompanying documents typically found in a code repository. Nevertheless, many repositories might lack READMEs or may include inadequately written documentation. This can lead to the presence of noisy signals within the representation, complicating the extraction of meaningful information.

The research question we address is whether repository-level representations can be effectively generated directly from source code, even without supervisory signals like READMEs. To summarize our findings and contributions:

\begin{itemize}
{
\item{\textbf{Introducing Topical}: We present ``Topical'', an attention-driven deep neural network architecture that  integrates information from three domains within a code repository:} 1) Dependencies 2) Code content 3) Docstrings.
This fusion is designed to generate holistic repository-level representations. Complementing Topical is a GitHub crawler we developed that collates repositories and associated featured topics. These topics serve as training data to refine Topical's repository-level embedding generation.}

\item{\textbf{Versatility of Topical}: Topical can be incorporated as a generalized, lightweight architecture for determining repository-level representations. This can be achieved by using any other pre-existing method-level representation, such as import2vec~\cite{theeten2019import2vec}.}

\item{\textbf{Benchmarking Topical's Performance}: Topical excels in the GitHub repositories auto-tagging task. It outperforms na\"ively aggregated repository-level representations derived from existing embeddings; Topical also has the edge over repository-level representations even when attention mechanisms are integrated with existing method-level representations.}
\end{itemize}

\subsection{Related Work}

Prior work into source code topic modeling have successfully adapted statistical methods that were previously employed for the topic modeling of textual documents within the domains of information retrieval (IR) and natural language processing (NLP)~\cite{2013genetics,chen2016survey,landauer1998introduction,Blei03,yi2009comparative}. Such techniques treat source code as a collection of independent tokens and apply statistical modeling using term frequency analysis methods, including TF-IDF (Term Frequency-Inverse Document Frequency) and Latent Dirichlet Allocation (LDA)~\cite{Blei03,Sharma17,hindle2009s,2013genetics,de2014labeling,chen2016survey}. The work of Ascuncion~\etal. illustrates the efficacy of LDA in surfacing topics within software artifacts, further emphasizing the importance of this approach in traceability tasks aimed at finding connections between related software entities~\cite{asuncion2010software}. Alternative methods include the automatic tagging of code with topics, utilizing ``README'' files from GitHub repositories~\cite{Sharma17}. To compare the performance of \emph{Topical} against conventional statistical approaches (such as LDA or TF-IDF), we introduce a specialized term frequency model TF3D. Analogous to Topical, TF3D capitalizes on terms' distributions across three repository domains, namely, source code, packages imports (dependencies), and docstrings. 

Recent advancements in the domain of deep learning have witnessed a shift from traditional term distribution models towards parametric language models, particularly models like BERT~\cite{devlin-etal-2019-bert, sanh2020distilbert}. Guo~\etal extended pre-trained BERT models to capture the nuances of programming languages, culminating in the development of CodeBERT~\cite{feng2020codebert}. This multilingual model is trained over a selection of programming languages. To augment CodeBERT's capabilities, its derivative, GraphCodeBERT, was devised to integrate code structural information, ensuring a more comprehensive representation. Furthermore, GraphCodeBERT~\cite{guo2021graphcodebert} has demonstrated superior performance, setting benchmark results on tasks associated with the CodeSearchNet dataset~\cite{husain2020codesearchnet}. In a parallel line of research, Theeten~\etal. introduced ``Code-Compass'', a novel tool leveraging their devised Import2Vec mechanism~\cite{theeten2019import2vec}. This mechanism represents script dependencies in a high-dimensional embedding space, where dependencies with inherent similarities naturally gravitate together. Within this space, each software package is mapped to a distinct vector, which Code-Compass subsequently employs to suggest related packages through similarity metrics. We use Import2Vec serves a baseline in this paper. Our empirical results show that combining Import2Vec with the Topical attention mechanism gives a performant repository-level representation. This combination outperforms the existing methods, such as mean aggregation or concatenation of script-level information~\cite{rokon2021repo2vec}.

Recent methodologies proposed for repository-level tagging predominantly rely on textual metadata, such as filenames, READMEs, and supplementary documentation present in code repositories~\cite{thung2012detecting,thung2013automated,chen2015simapp,zhang2017detecting}. Izadi~\etal's research exemplifies this trend, where repository topics are derived by analysing filenames, READMEs, and associated Wiki data~\cite{izadi2021topic}. Implementing DistilBERT on tokens, they generated a repository embedding. Subsequently, a fully-connected layer with a sigmoid activation function was used for multi-label tagging. In contrast, our approach Topical, extracts  information exclusively from script-level data (\ie, code) and intentionally omits auxiliary textual content like READMEs or Wiki data. The extraction of useful detail from READMEs rests upon the quality of the document, authored by the developer. Such dependency can skew the representation, rendering it susceptible to misinformation.

Rokon~\etal propose Repo2vec~\cite{rokon2021repo2vec}, a tool for producing repository-level embeddings. It integrates documentation, including READMEs, metadata, and source code into its input. Specifically, Repo2Vec combines insights from source code, READMEs, and the repository structure, consolidating these modalities into a unified embedding by vector concatenation. This framework is then calibrated on a dataset comprising human-assessed labels of repository similarities. In contrast, Topical is trained on a dataset that is automatically generated, negating the necessity for manual annotations – a process that may be challenging to scale and objectively evaluate. Topical solely leverages the ``content'' - encompassing the source code, its structure, and related docstrings. Incorporating an attention mechanism, our empirical results underscore Topical's superior classification ability over embeddings produced through conventional aggregation techniques like mean or concatenation. Preliminary observations show that the attention mechanism can surpass Repo2vec's performance in similarity assessments. We place an emphasis on repositories without exhaustive README documentation or those without of any documentation, primarily encoding the source code and its structure into a dense embedding. Our automated data acquisition is not solely restricted to those repositories with extensive documentation.

\begin{figure}[h!]
\centering
\subfloat[]{\includegraphics[scale=0.22]{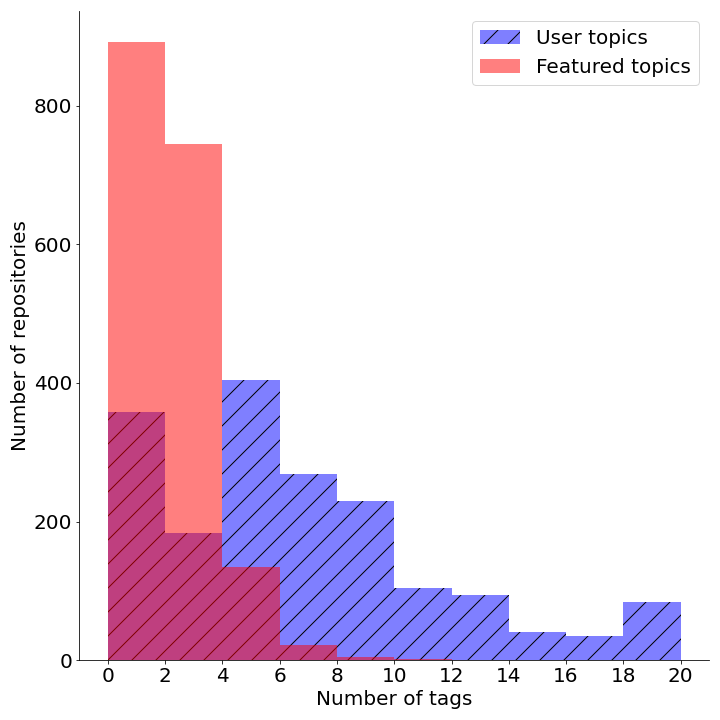}}\quad

\subfloat[]{\includegraphics[scale=0.44]{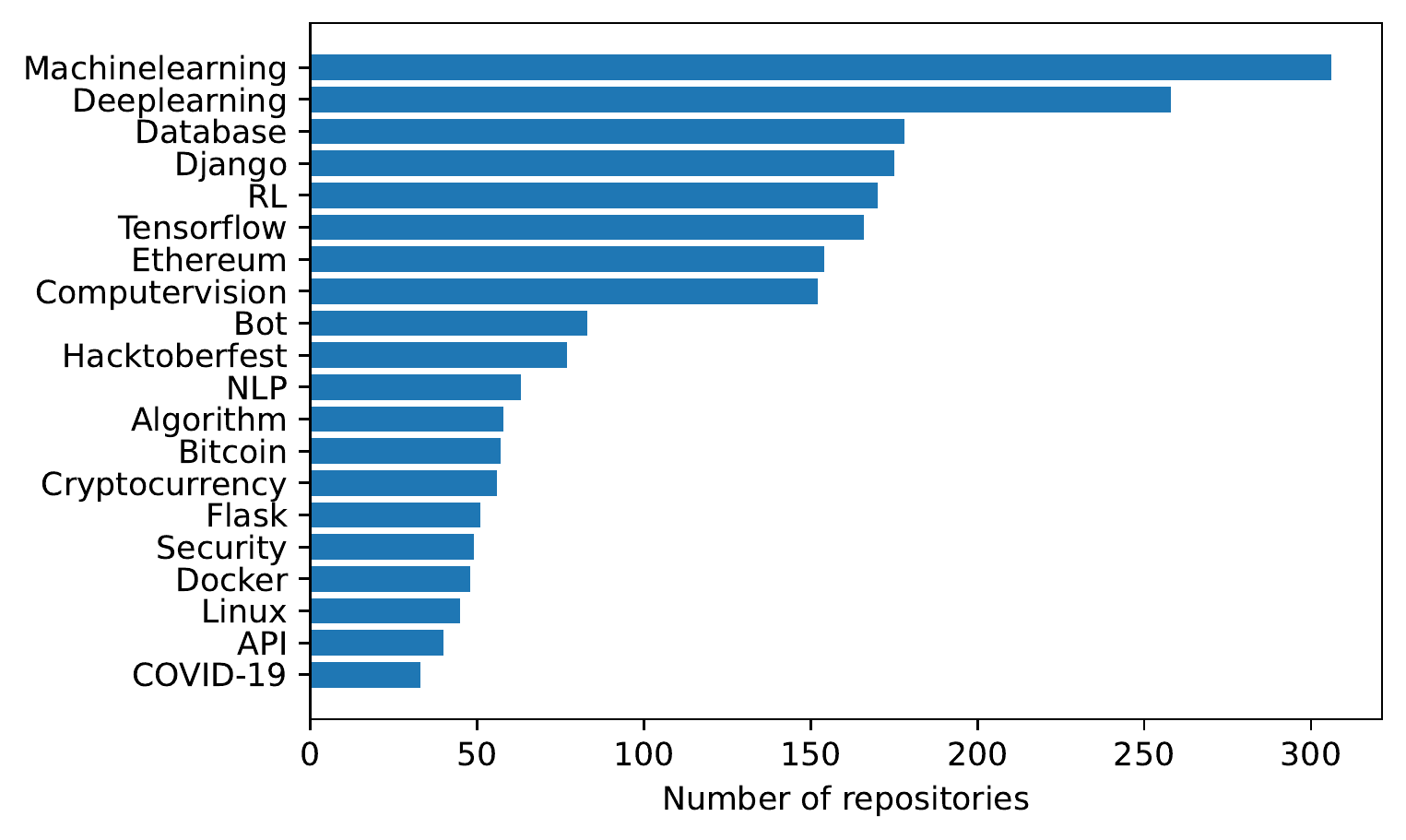}}
\caption{(a) Distribution of the number of occurrences for the top 20 featured topics (red), user topics (blue), before and after Fuzzy Matching (mixed). (b) Distribution of repositories by topics in the dataset.}
\label{fig:featured_topics}
\end{figure}

\begin{figure*}[h!]
\centering
\subfloat[]{\includegraphics[scale=0.18]{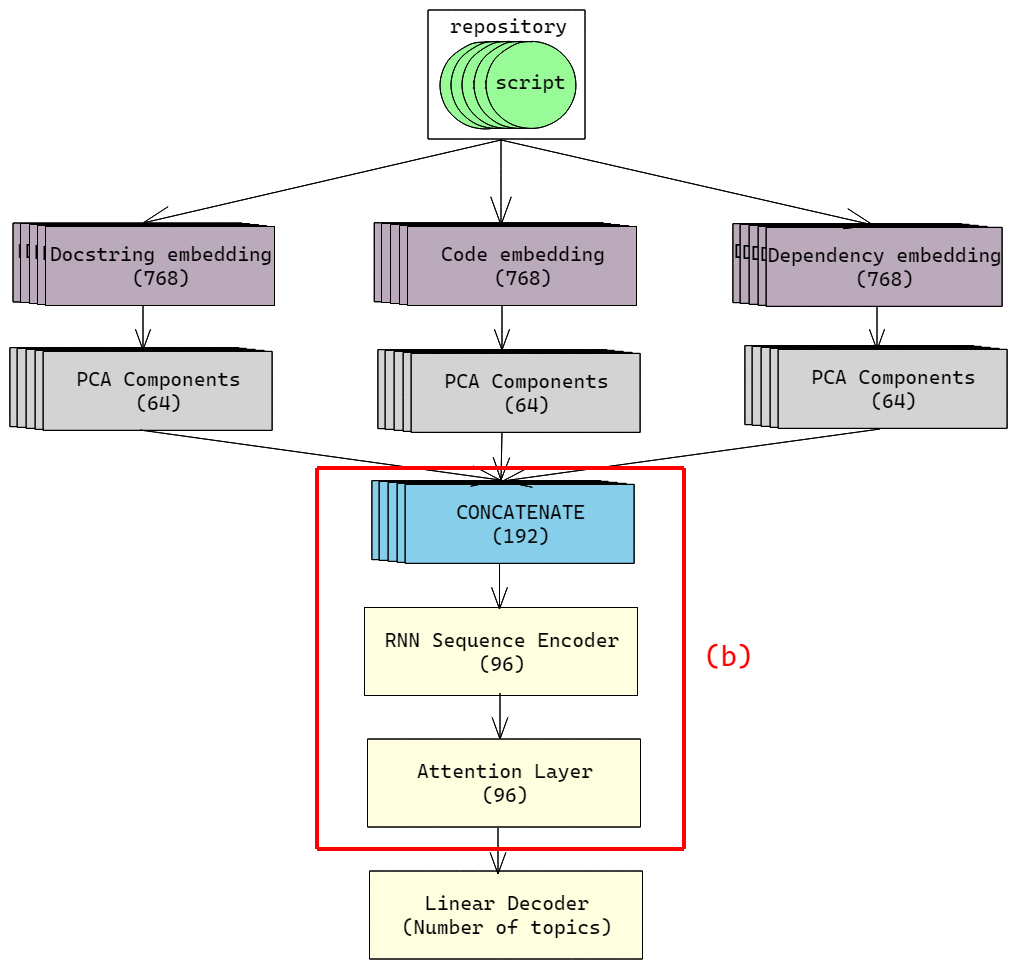}}
\subfloat[]{\includegraphics[scale=0.34]{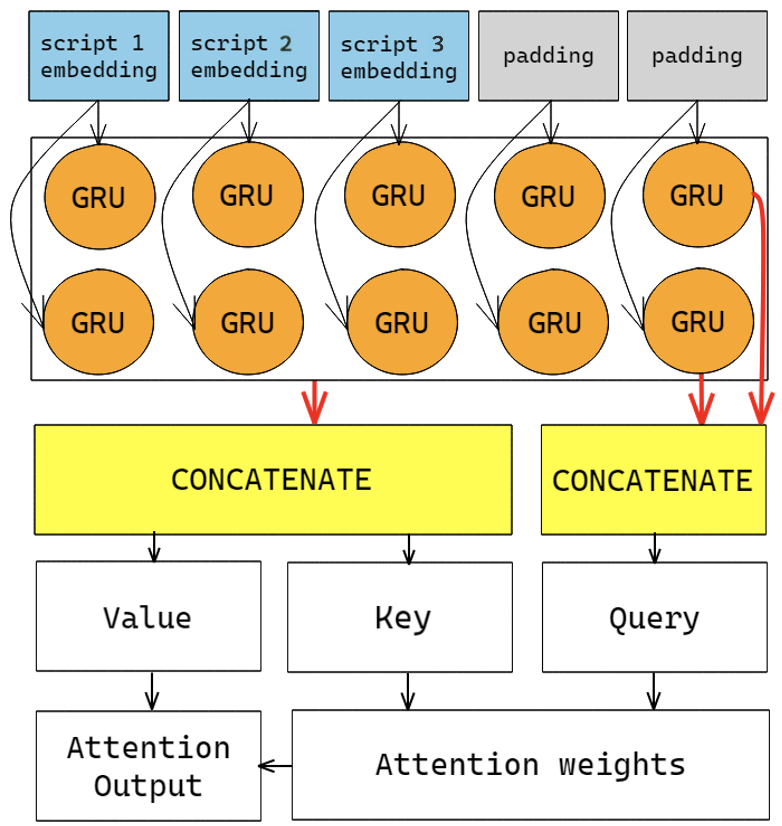}}
\caption{(a) General model diagram including downstream classification layer. (b) RNN sequence encoder on script embeddings with Attention layer to compute a single embedding representation of the repository.}
\label{fig:model_scheme}
\end{figure*}

\section{Methodology}\label{sec:methods}
\subsection{Dataset and GitHub Crawler}

We create a dataset at the repository level, complemented by associated topic annotations for the respective repositories. Predominant open-source code datasets, such as those described by Husain~\etal~\cite{husain2020codesearchnet} and Puri~\etal~\cite{puri21codenet}, are primarily tailored for file or method-level tasks. These datasets lack comprehensive details necessary for analyzing aggregations of files, including import dependencies, metadata, and commit/git history. We focus exclusively on repositories written in Python, though Topical is adaptable to other programming languages. No benchmark datasets of repositories with corresponding annotations exist that align with the objectives of this study.

To bridge this gap and contribute to forthcoming research on source code topic modeling, we have developed a GitHub crawler tool. This tool is engineered to extract datasets from open-source repositories. For data acquisition, we used a predetermined set of 20 topics (\eg, Machine Learning, Natural Language Processing, Database management, \etc.). The crawler was programmed to retrieve a set number of repositories linked to each specified topic. The culmination of this extraction yielded a dataset of approximately 3,000 repositories, encapsulating an estimated 92,383 Python scripts in total. Notably, about 32\% of these repositories lacked a designated featured topic. Considering the multiplicity of featured topics that a single repository may possess, the resultant featured topic distribution exhibited a bias towards specific subjects. For the purpose of the classification task, repositories corresponding to the top 20 most recurrent featured topics were selected, resulting in a dataset of 1,600 repositories. The final distribution of topics is shown in Fig.\ref{fig:featured_topics}(b).

\subsection{Model}
Our primary objective is to formulate a representation of source code at the repository level. We developed Topical, a model that integrates a set of encoders combined with a deep attention mechanism. Topical is illustrated in Figure~\ref{fig:model_scheme}. The model has three distinct phases. Firstly, the model generates embeddings from scripts within a repository. To achieve this, we use pre-trained BERT base transformer model, generating embedding vectors for every script within a repository. Three embedding vectors are produced for each script, which is described in section~\ref{sec:ScriptlevelEmbedding}. Following the embedding phase, the model applies an attention mechanism, which combines individual script embeddings into a unified representation. Finally, the architecture incorporates a classifier unit for classification of repository topics. Each of these stages are described in subsequent sections and in the supplementary material.

\subsection{Script-level Embedding}\label{sec:ScriptlevelEmbedding}

Topical integrates three domains within a repository as input vectors to its encoder. These domains encompass the content and structure of the source code, the embedded textual information within the source code, including metadata, and the dependency graph. The latter is constructed based on method calls within scripts, calls directed towards specific classes, and calls to external libraries. While the content and associated textual metadata of the source code, such as documentation strings and file names, are conventionally employed in code classification as underscored by Guo~\etal~\cite{guo2021graphcodebert}, the dependency graph domain serves a broader purpose. Beyond enhancing topic tagging accuracy, it aids in the identification of recurrent patterns within software architectures, a perspective highlighted by Theeten~\etal~\cite{theeten2019import2vec}. In more detail: (i) \textbf{Embedding Code Content and Structure}
The content of the code is embedded through the GraphCodeBERT RoBERTa base model~\cite{guo2021graphcodebert} which has been pre-trained on multiple coding languages. For our purposes, we set the input token size at 512 for every script, consistent with the GraphCodeBERT default settings. (ii) \textbf{Textual Information: Embedding Docstrings}
 File names, method names, and docstrings are concatenated, tokenized, and separated using distinct tokens. The resulting tokenized data is then encoded employing the DistilBERT model, maintaining an input token size of 512. (iii) \textbf{Embedding the Dependency Graph}: existing methods have leveraged library import declarations to give embeddings of imported library dependencies~\cite{theeten2019import2vec}, representing these as abstract trees on the repository scale. Yet, these methodologies predominantly rely on package loading instructions, which might not accurately reflect code communication and real-time library application. We propose an embedding technique that encapsulates the communication graph between methods and scripts. To achieve this, we utilize PyCG~\cite{salis2021pycg}, an open-source tool, designed to retrieve a dependency graph from static Python code. Given the inherently descriptive nature of methods and package denominations, we leverage a DistilBERT model, which has been pre-trained on the English language, to facilitate the embedding of the graph. The supplementary material provides a discussion on this.

\subsection{Repository-level Embedding}
Topical employs an attention mechanism to combine embeddings derived from distinct pre-trained BERT models (namely GraphCodeBERT and DistilBERT) into a unified embedding for each individual script within a given repository. A schematic of the model's architecture is given in Fig.~\ref{fig:model_scheme}a-b. To enhance computational efficiency and ensure the model's scalability to large datasets, we reduce the number of components derived from each embedding that feeds into the final script representation. This lowers computational demand and also reduces the number of parameters, facilitating optimization tailored to the specific downstream task. For the purpose of dimensionality reduction, we apply PCA to transform the 768-dimensional embedding vectors into a condensed space of 192 dimensions. After dimensionality reduction, embeddings from various scripts are combined into a single, dense repository representation using attention. This is achieved by an encoder augmented by a self-attention mechanism~\cite{Vaswani17}. We explore different encoder architectures in our experiments: bi-GRU~\cite{chung2014empirical}, MLP, and bi-LSTMs. 

The self-attention layer allows Topical to optimize the distribution of weights for the scripts. This is useful for classifying topics as the topic can manifest in a small fraction of the repository or by distinctive relationship between topics in a collection of scripts.
Our attention paradigm for a bi-GRU sequence encoder is detailed in Fig.~\ref{fig:model_scheme}b.
For a given repository, $R$, we obtain $R^d=\lbrace x_0, x_1,\dotsc,x_n\rbrace$, as the script embedding for each domain $d$, where $d$ belongs to either code structure, docstrings, or dependencies. $x_t$ is therefore a single obtained script embedding, where $t$ is the script position in a sequence of scripts from a repository which is used as an input to the GRU hidden layer. 
Since we do not consider here the order of the scripts sequence, we utilise attention on the bi-directional GRUs. The hidden layer of the forward GRU is represented as $\overrightarrow{h_t}$, and the hidden layer of the backward GRU is represented as  $\overleftarrow{h_t}$, where we calculate:
\begin{equation}
    \overrightarrow{h_t} = GRU(x_t, \overrightarrow{h_{t-1}})
\end{equation}
\begin{equation}
    \overleftarrow{h_t} = GRU(x_t, \overleftarrow{h_{t-1}}) 
\end{equation}

\noindent{For a repository composed of $n$ scripts, we retrieve the last hidden state of both hidden layers and concatenate them as follows:}

\begin{equation}
    h_n = [\overrightarrow{h_n}, \overleftarrow{h_n}]
\end{equation}
We also retrieve the output of the GRU which is the tensor of all its hidden states:
\begin{equation}
    y = [h_i]_{0\leq i \leq n},  h_i = [\overrightarrow{h_i}, \overleftarrow{h_i}]
\end{equation}
\noindent{Indeed, $h_n$ contains information from all the other hidden states and thus permits to represent the entire collection of scripts. $y$ is then used as the key and value to the attention layer while the last hidden state $h_n$ is used as query.}

\subsubsection{Sampling scripts from repository}

For $y$ to maintain a uniform structure, when the cumulative number of scripts falls short of $n$, we introduce padding embeddings (\eg~script embeddings comprised entirely of zeros) to the sequence. In this study, we examine the implications of adjusting the cap on script files, with $n \in \{2,5,10,15\}$, intended for embedding within an individual repository. While repositories can house a vast number of scripts, setting a restricted count facilitates the representation of repositories that are still undergoing development. Direct random sampling from sizable repositories can lead to an overrepresentation of files from replicated third-party libraries. To address this, we deploy PyCG to curate scripts for the embedding procedure. This method involves tracing script paths from the repository's primary directory, focusing on those scripts that participate in the function call sequence. Scripts are integrated based on their sequence within the pathway. Upon completion of a given pathway, a new path is randomly selected and explored to fulfill the quota of $n$ scripts.

We apply an attention mask onto these padding embeddings when computing the attention output~\cite{Vaswani17}.
The mask matrix computation sets attention weights to 0 on padding embeddings using attention as:
\begin{equation}
    F = softmax(\frac{Q \times k^T + M}{\sqrt{d_k}}) \times V
\end{equation}
where $M$ the mask matrix is:
\begin{equation}
M_{t, i} = 
\begin{cases}
0 & \text{if } x_t \text{ is a script embedding} \\
- \infty & \text{if } x_t \text{ is a padding embedding}
\end{cases}
\end{equation}
and $Q$ is the query matrix, $k$ the key vector, $V$ the value matrix as shown in Fig.~\ref{fig:model_scheme}(b) and $d_k$ is the dimension of the key vector.

\subsubsection{Multi-label Classification}
Most of the repositories available on GitHub belong to more than one topic. Furthermore, some featured topics are subtopics of others (Fig. \ref{fig:featured_topics}b). For example, an NLP focused repository will likely be assigned to the broader topic machine learning. We picked 5-20 representative topics by their frequency that can also be found separate to other topics. To enable the multi-label classification task, we add a linear layer on top of the attention mechanism, paired with a sigmoid activation function. The architecture is trained to minimize the cross entropy loss between the predictions and the ground truth label. During inference, given output scores for each class $s_i$, we predict labels using thresholding,~\ie.
\begin{equation}
    \hat{l}_i = \left\{
    \begin{array}{ll}
    1 &\mbox{if } s_i\geq threshold \\
    0 &\mbox{else}
    \end{array}
    \right.
\end{equation}
where the threshold is chosen to maximize the $F_{1}$ score for topic $i$ on a validation set. 

\subsection{Baselines} 
In order to provide conclusive results for our attention based model, we develop multiple competitive baseline models, TF3D, GraphCodeBert, and Import2vec. TF3D is a novel model we have developed that is based on term frequency. This model allows us to compare Topical to a non deep learning model, but with similar embedding information (source code, docstrings, and dependencies). The second baseline, GraphCodeBERT uses embedding based only on code content of the repository. Finally, we use the repository embedding model, Import2vec, in four new variations which are built on top of the pre-trained Import2vec model. The performance of these baselines highlights the flexibility and effectiveness of Topical when combined with existing script/file level embedding.

\subsubsection{\textbf{TF3D - A Statistical NLP Baseline}}

TF3D is based on representing terms frequencies, such as TF-IDF, but adapted in this case to collection of scripts/methods. Typical statistical term-based models for source code were recently shown to be effective for source code analysis and similarity detection~\cite{tf_azcona2019user2code2vec,tf_fu2017wastk,tf_islam2020socer}. Similar to the attention model we combine three source-code feature domains: (a) the code structure, by using AST (Abstract Syntax tree) features of each method in code;  (b) The \emph{docstrings}, which are any comments at the method level, and function names and finally, (c) dependency/libraries of script files. We represent a repository as a collection of $n$ methods with their corresponding feature vectors ($m$) in the source code of a repository, $i$, $R^d_i=\lbrace m_{1}, m_{2}, \dotsc, m_{n}\rbrace$ where $d$ belongs to either code structure, docstring or dependencies. For each repository in each $d$, $R^d_i$, we use aggregation to represent the probability vector of features in a repository:
\begin{equation}
    S_i^d=\frac{\sum_j{m_{j}}}{\|m\|}
\end{equation}
By sampling a training set of $N$ repositories $\lbrace R_0, R_1,\dotsc, R_{N}\rbrace$ we calculate the terms matrix, $C(3 \times n_T)$, for each topic ($T$) in $n_T$ topics by using the arithmetic mean on the logarithm of $S^d_i$:
\begin{equation}
    C(d,T)=\frac{\langle\ln{S_{i\in T}}\rangle}{\langle\ln{S}_{i\notin T}\rangle}
    \label{eq:clarity}
\end{equation}

Note that here we modified the standard TF-IDF~\cite{jones1972statistical} and introduce the logarithm to penalise excessively repeating terms in scripts or methods in $R_i$, that can dominate the frequency vectors (instead of penalising the inverse of the frequency as typically used). 
Then, we calculate the cosine similarity, to obtain the embedding matrix representing each repository, $i$, in the training and testing sets: $E_i(3 \times n_T) =\frac{S_i\cdot C^T}{\|C\|\|S_i\|}$. Finally,  we use the embedding matrices for classification using a standard random forest regressor classifier for multi-label tagging of repositories.

\subsubsection{\textbf{GraphCodeBERT}}
GraphCodeBERT by itself combines code-content, its \emph{data-flow}, and comments found within methods. GraphCodeBERT embeds method level source code for various code-related tasks. We therefore use it as a baseline to show the efficiency of our attention model. 

We have experimented with other models for method level source code embedding, such as CodeBERT infused with code AST using Attention mask. In that model, CodeBERT relies only on source code content and using AST graph connections as attention masks enables to capture code data flow. However, the GraphCodeBERT provided significantly better results, and we therefore report only GraphCodeBERT as a competitive baseline that uses pre-trained method level embedding to represent a repository. Compared to Topical that combines the docstring, code and Dependency embedding, GraphCodeBERT baseline only uses the code embedding. 

\subsubsection{\textbf{Import2Vec}}
The final comparison is made against a model utilizing the Import2vec embedding, as outlined by Theeten~\etal (2019)~\cite{theeten2019import2vec}. This comparison further demonstrates the compatibility of the Topical architecture with existing method-based embeddings for the generation of repository-level embeddings. Import2vec offers vector representations for software libraries, based on their incorporation within scripts. The basic principle of this embedding is predicated upon the semantic affinity between libraries, specifically, the likelihood of their concurrent importation. Theeten~\etal.~\cite{theeten2019import2vec} provide pre-trained Import2vec embeddings, yielding vector representations for a myriad of extant Python libraries/packages, such as numpy, tensorflow, among others. For leveraging the Import2vec embeddings in downstream tasks, particularly classification in our context, we start by extracting an exhaustive list of software libraries imported within a repository. For every library integrated into the repository, its corresponding Import2vec vector is computed. These vectors have a dimensionality spanning from 60 to 200. As suggested by Theeten et al.~\cite{theeten2019import2vec}, an embedding dimension of 100 is effective. For classification, a basic strategy involves either computing the mean of these vectors or concatenating them in a structured order, which then feeds into a classifier. We combine these vectors using the Topical attention mechanism. This translates to representing each repository as an array of Import2vec vectors $R = \{x_0, x_1, \dots, x_n \}$ corresponding to the libraries it integrates. We obtain four variations of the Import2vec embedding as baselines: 

\begin{itemize}
\item \textbf{I2V-conc-linear:} Concatenation of the Import2vec embeddings associated with all libraries integrated by a repository in a pre-determined sequence, followed by the application of a linear layer for repository auto-tagging.
\item \textbf{I2V-conc-attn:} Concatenation of the Import2vec embedding in same way as i2v-conc-linear, however, these embedding vectors are then combined using the Topical attention mechanism architecture as shown in Fig.~\ref{fig:model_scheme}(b).
\item \textbf{I2V-mean-linear:} The average of the Import2vec embedding vectors of all the libraries imported by a repository followed with a linear layer
\item \textbf{I2V-mean-attn:}  The average of the Import2vec embedding vectors of all libraries imported by a repository and then processed by the Topical attention mechanism.
\end{itemize}

\subsubsection{Evaluation Metrics}

\begin{itemize}
    \item \emph{LRAP (Label Ranking Average Precision)}: The LRAP is calculated as follows:
\begin{equation}
LRAP(y, \hat{f}) = \frac{1}{n_{samples}}\sum^{n_{samples}-1}_{i=0}\frac{1}{||y_i||_0}\sum_{j:y_{ij}=1}\frac{|\mathcal{L}_{ij}|}{\text{rank}_{ij}}
\end{equation}
\[\mathcal{L}_{ij} = \{k : y_{ik} = 1, \hat{f}_{ik} \geq \hat{f}_{ij}\}\]
\[\text{rank}_{ij} = |\{k : \hat{f}_{ik} \geq \hat{f}_{ij}\}|
\]
where $y \in \{0,1\}^{n_{samples} \times n_{labels}}$ is the binary indicator matrix of the ground truth labels and $\hat{f} \in R^{n_{samples}\times n_{labels}}$ is the vector of scores predicted for each label. The $\text{rank}_{ij}$ provides the index of the ordered prediction vector, and $|\mathcal{L}_{ij}|$ is the number of true predictions for all indices above $\text{rank}_{ij}$. Thus, the average over the ratio for all samples gives a reliable metric for multi-label scores. LRAP measures the quality of a multi-label classification by first ranking the labels predicted by a model and then reasoning about order in which the correct labels appear. If the correct labels appear at top ranks then a score of 1 (or close to 1) is awarded, however if correct labels appear at lower ranks then, depending on their rank, only a small fraction of 1 is awarded. 

\item \emph{$F_{1}$ scores}: In addition to LRAP, we report the micro-average $F_{1}$ score \cite{daume2017course} computed by counting the global true positive, false positive and false negative across the dataset.

\end{itemize}

\begin{table*}[h]
  \centering
  \caption{Comparison table for Topical model with mean-GraphCodeBERT and TF3D baselines (Validation set).}
   \label{tab:precrec}
   \resizebox{1.2\columnwidth}{!}{%
   \begin{tabular}{lccc}
   \hline
      \textbf{Model} & \textbf{Precision} & \textbf{Recall} & \textbf{Optimized Threshold} \\
      \hline
      Topical & \textbf{0.485 $\pm$ 0.017} & 0.630 $\pm$ 0.032 & 0.217 $\pm$ 0.025 \\
   GraphCodeBERT &0.410 $\pm$ 0.031 & \textbf{0.670 $\pm$ 0.010} & 0.140 $\pm$ 0.011 \\
     I2V-conc-attn & 0.350 $\pm$ 0.034 & 0.632 $\pm$ 0.034 & 0.187 $\pm$ 0.011 \\
    \hline
   \end{tabular}
}\end{table*}

\section{Results}\label{sec:results}
\begin{figure}[h!]
\centering
\subfloat[]{\includegraphics[scale=0.264]{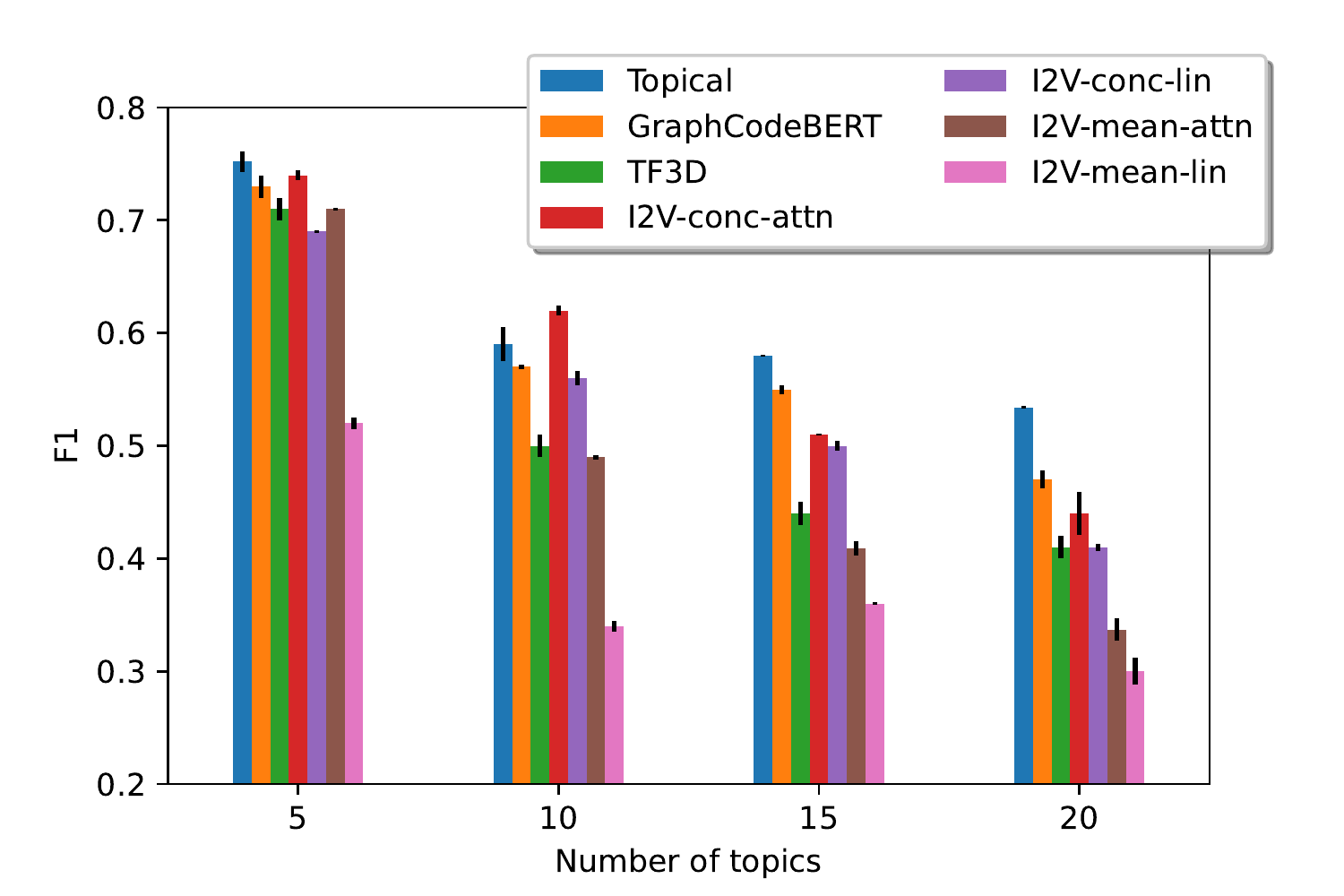}}\quad
\subfloat[]{\includegraphics[scale=0.264]{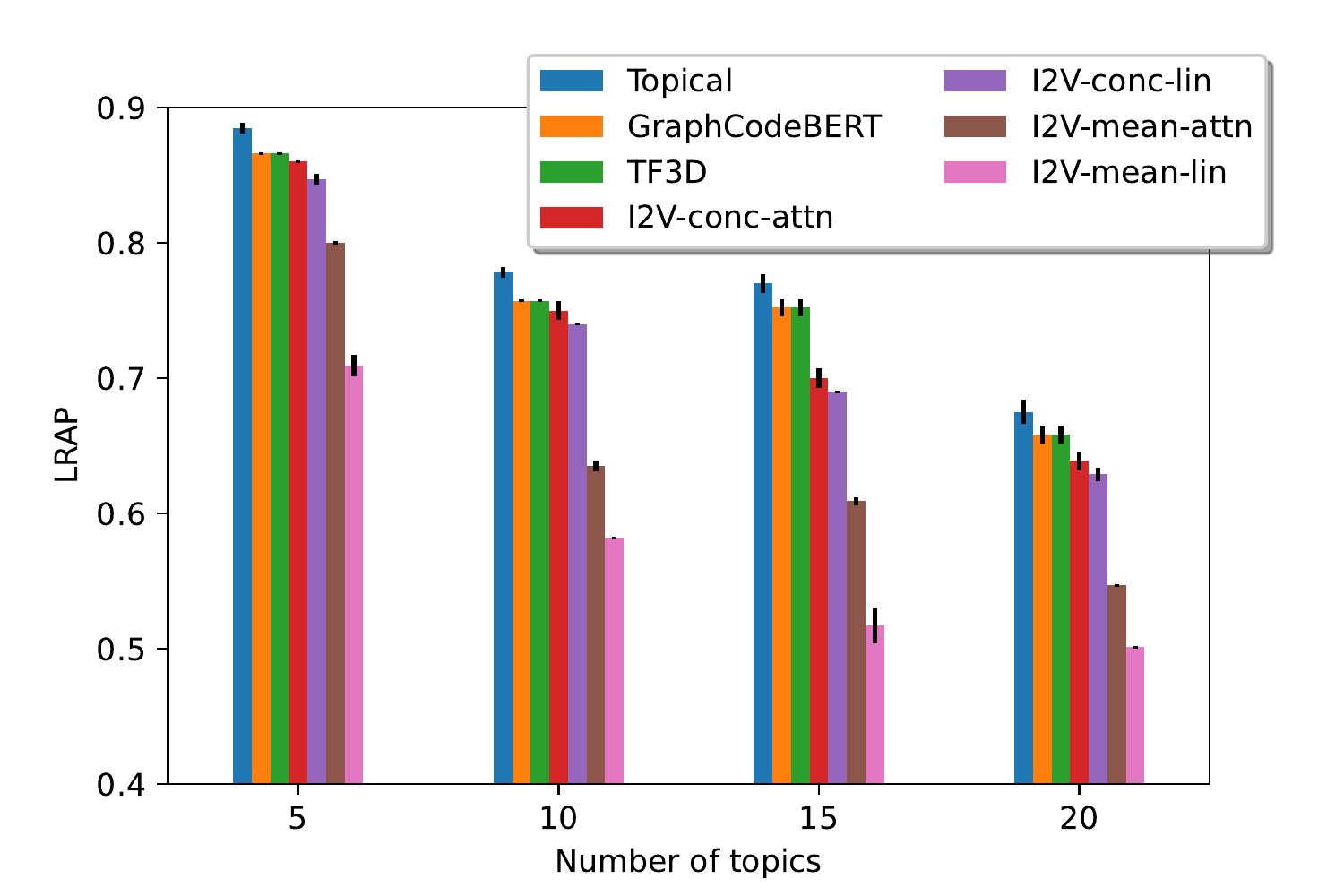}}
\caption{(a) $F_{1}$-score comparison and (b) LRAP score comparison for various baselines for auto-tagging for 5,10,15 and 20 topics.}
\label{fig:topics}
\end{figure}

We compare the performance of various baselines for auto-tagging on different number of topics ($\{5,10,15,20\}$). We also compare the performance of various algorithms across the size of dataset to compare how well they scale with data. All algorithms are trained by minimising the cross-entropy loss with a fixed learning rate of 0.002 using ADAM optimiser with weight decay \cite{loshchilov2017decoupled, kingma2014adam}. Classification results for Topical in comparison to the baselines are shown in Figure \ref{fig:topics}. Topical shows overall better $F_{1}$-score than TF3D and GraphCodeBERT baselines, and 6\%-10\% higher performance for LRAP scores. This is also true for $F_{1}$-scores as Topical does better than the rest of the baselines. Overall, Import2vec based baselines perform slightly poorer than GraphCodeBERT and TF3D indicating that processing the full source code has significant advantage over only looking at software libraries imported by a repository. 
A second insight is that combining pre-trained embeddings with the attention mechanism can result in significant performance gain. For example, the performance of Import2vec embedding grows  when using the Topical attention as compared to na\"ive aggregation mechanisms such as concatenation and averaging. Another trend revealed by the results of various topics number, is that as the number of topics increases the performance starts to drop. However, one likely reason for that could be the overlap between similar topics. 

\vspace{-2mm}
\subsubsection{Effect of varying data size}
\begin{figure}[h!]
\centering
\subfloat[]{\includegraphics[scale=0.26]{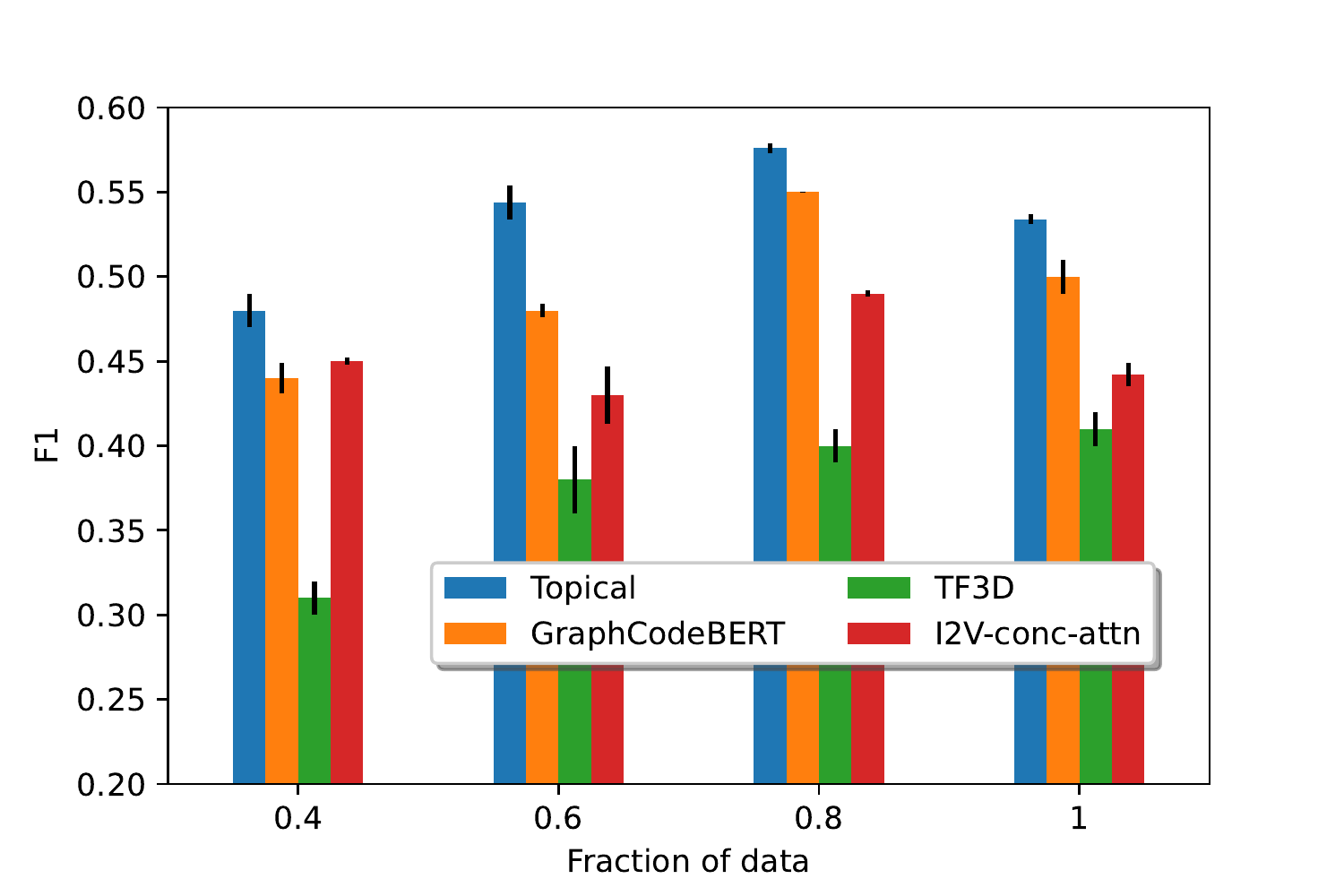}}\quad
\subfloat[]{\includegraphics[scale=0.26]{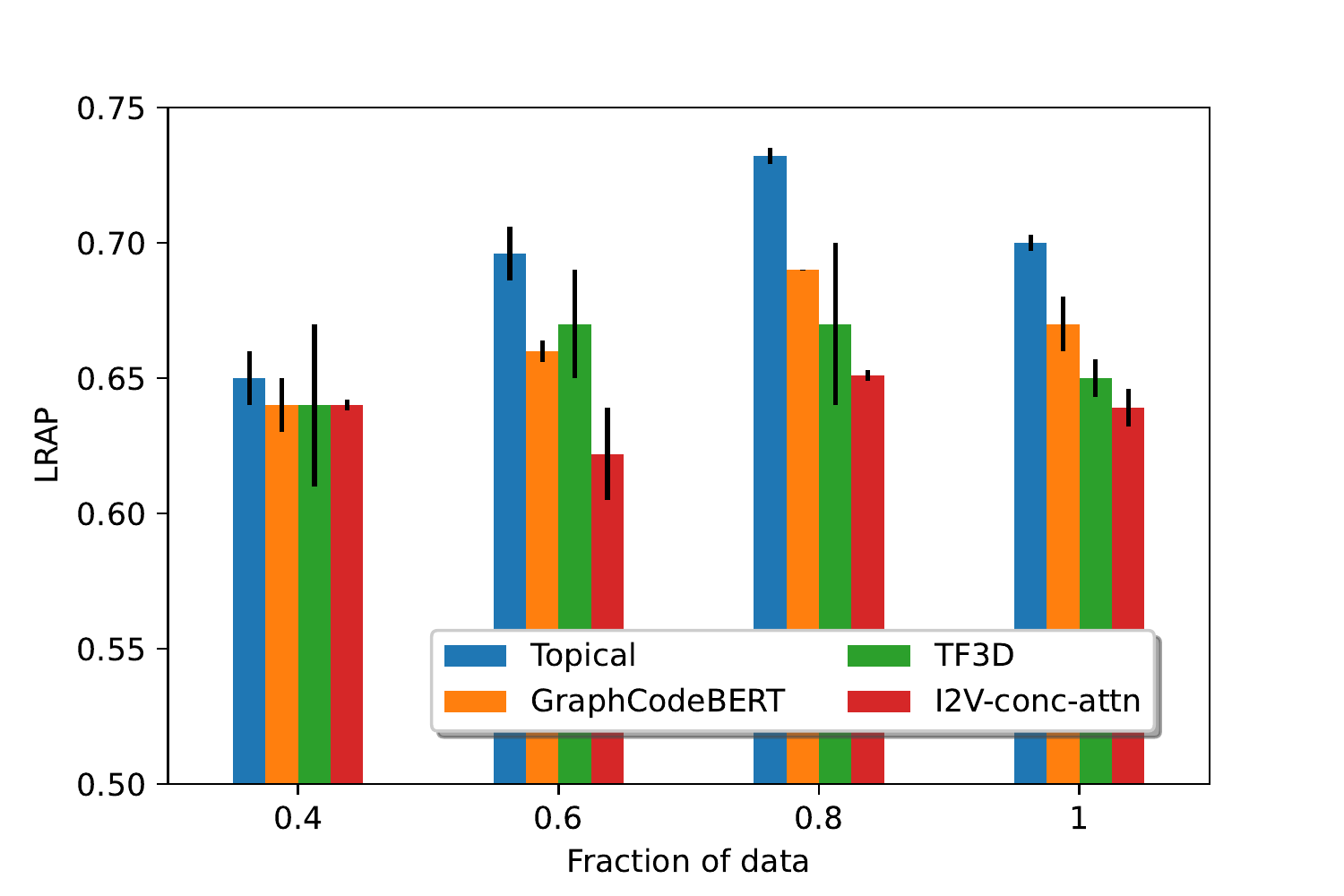}}
\caption{(a) $F_{1}$-score comparison for 20 topics multi-label classification on 40\%, 60\%, 80\% and full dataset size (b) LRAP score comparison for 20 topics multi-label classification on 40\%, 60\%, 80\% and full dataset size.}
\label{fig:size}
\end{figure}

In Figure \ref{fig:size}, the $F_{1}$-score and LRAP are presented in relation to the fraction of the training set for 20-topic classification. Using a deep embedding attention model, we find that a modest number of repositories can yield satisfactory performance. This indicates the potential for efficiency in both training and data acquisition, even when targeting new topics or different classification tasks. As hypothesized, TF3D performs optimally when data is limited. But as the size of the training set expands, the attention model, anchored by deep repository embedding, tends to exhibit higher metrics. With an increase in dataset size, there's a gradual improvement in the performance metrics across various models, eventually approaching a point of saturation.

\subsection{Discussion}
We can summarise the outcome of our study as follows:
\begin{itemize}
\item \textbf{How can we extract extract a useful representation from different sources of information in a code repository? }
Topical extracts and combines method-level representations and dependencies graph to compute repository level representations. Topical outperforms baselines that include representations based on only source code information.
\item \textbf{How can we combine method level representations to generate repository level embedding?} Topical can be used to generate repository level representation from existing method level representations (import2vec). The repository level representation computed by Topical outperforms other aggregation methods.
\end{itemize}

Table~\ref{tab:precrec} provides metrics for three attention-based baselines. For a 20 class multi-label classification task, all three algorithms display adequate recall values. However, precision seems to differentiate their performances. Topical, while not having the maximum recall among the baselines, has higher precision. Coupled with the LRAP scores, this suggests that Topical assigns topics accurately and gives them higher ranks. Conversely, the GraphCodeBERT and Import2Vec baselines tend to select a wider range of topics and rank them lower, leading to their lower precision and LRAP scores, but maintaining recall. It's also observed that for 20 topics, Topical and the other attention-based baselines produce an LRAP score above 0.6. This indicates that these baselines often rank relevant topics within the top 50\% of ranks, suggesting a potential avenue for optimization by focusing on higher-ranked topics.

\section{Conclusions}\label{sec:conclusions}
We introduce Topical, a tool designed to generate repository level embeddings from source code using GitHub's curated topics. We find that the attention mechanism is effective in improving script level embedding for subsequent tasks. The tool utilizes pre-trained BERT models, making it cost-effective, transferable, and hardware-independent for various tasks and topics. Future work will focus on employing Topical embeddings to summarize repositories in natural language.

\bibliography{aaai24}

\begin{thebibliography}{49}
\providecommand{\natexlab}[1]{#1}

\bibitem[{Allamanis et~al.(2018)Allamanis, Barr, Devanbu, and
  Sutton}]{Allamanis18}
Allamanis, M.; Barr, E.~T.; Devanbu, P.; and Sutton, C. 2018.
\newblock A Survey of Machine Learning for Big Code and Naturalness.
\newblock \emph{ACM Comput. Surv.}, 51(4).

\bibitem[{Asuncion, Asuncion, and Taylor(2010)}]{asuncion2010software}
Asuncion, H.~U.; Asuncion, A.~U.; and Taylor, R.~N. 2010.
\newblock Software traceability with topic modeling.
\newblock In \emph{2010 ACM/IEEE 32nd International Conference on Software
  Engineering}, volume~1, 95--104. IEEE.

\bibitem[{Azcona et~al.(2019)Azcona, Arora, Hsiao, and
  Smeaton}]{tf_azcona2019user2code2vec}
Azcona, D.; Arora, P.; Hsiao, I.-H.; and Smeaton, A. 2019.
\newblock user2code2vec: Embeddings for profiling students based on
  distributional representations of source code.
\newblock In \emph{Proceedings of the 9th International Conference on Learning
  Analytics \& Knowledge}, 86--95.

\bibitem[{Blei, Ng, and Jordan(2003)}]{Blei03}
Blei, D.~M.; Ng, A.~Y.; and Jordan, M.~I. 2003.
\newblock Latent Dirichlet Allocation.
\newblock \emph{J. Mach. Learn. Res.}, 3(null): 993–1022.

\bibitem[{Buratti et~al.(2020)Buratti, Pujar, Bornea, McCarley, Zheng,
  Rossiello, Morari, Laredo, Thost, Zhuang, and Domeniconi}]{Buratti20}
Buratti, L.; Pujar, S.; Bornea, M.~A.; McCarley, J.~S.; Zheng, Y.; Rossiello,
  G.; Morari, A.; Laredo, J.; Thost, V.; Zhuang, Y.; and Domeniconi, G. 2020.
\newblock Exploring Software Naturalness through Neural Language Models.
\newblock \emph{CoRR}, abs/2006.12641.

\bibitem[{Chen et~al.(2015)Chen, Hoi, Li, and Xiao}]{chen2015simapp}
Chen, N.; Hoi, S.~C.; Li, S.; and Xiao, X. 2015.
\newblock SimApp: A framework for detecting similar mobile applications by
  online kernel learning.
\newblock In \emph{Proceedings of the eighth ACM international conference on
  web search and data mining}, 305--314.

\bibitem[{Chen, Thomas, and Hassan(2016)}]{chen2016survey}
Chen, T.-H.; Thomas, S.~W.; and Hassan, A.~E. 2016.
\newblock A survey on the use of topic models when mining software
  repositories.
\newblock \emph{Empirical Software Engineering}, 21(5): 1843--1919.

\bibitem[{Chung et~al.(2014)Chung, Gulcehre, Cho, and
  Bengio}]{chung2014empirical}
Chung, J.; Gulcehre, C.; Cho, K.; and Bengio, Y. 2014.
\newblock Empirical evaluation of gated recurrent neural networks on sequence
  modeling.
\newblock \emph{arXiv preprint arXiv:1412.3555}.

\bibitem[{Cohen(2014)}]{cohen2011fuzzywuzzy}
Cohen, A. 2014.
\newblock FuzzyWuzzy: Fuzzy String Matching in Python, July 2011.
\newblock \emph{URL http://chairnerd. seatgeek.
  com/fuzzywuzzy-fuzzy-string-matching-in-python/}, 1.

\bibitem[{Daum{\'e}(2017)}]{daume2017course}
Daum{\'e}, H. 2017.
\newblock \emph{A course in machine learning}.
\newblock Hal Daum{\'e} III.

\bibitem[{De~Lucia et~al.(2014)De~Lucia, Di~Penta, Oliveto, Panichella, and
  Panichella}]{de2014labeling}
De~Lucia, A.; Di~Penta, M.; Oliveto, R.; Panichella, A.; and Panichella, S.
  2014.
\newblock Labeling source code with information retrieval methods: an empirical
  study.
\newblock \emph{Empirical Software Engineering}, 19(5): 1383--1420.

\bibitem[{Devlin et~al.(2019)Devlin, Chang, Lee, and
  Toutanova}]{devlin-etal-2019-bert}
Devlin, J.; Chang, M.-W.; Lee, K.; and Toutanova, K. 2019.
\newblock {BERT}: Pre-training of Deep Bidirectional Transformers for Language
  Understanding.
\newblock In \emph{Proceedings of the 2019 Conference of the North {A}merican
  Chapter of the Association for Computational Linguistics: Human Language
  Technologies, Volume 1 (Long and Short Papers)}, 4171--4186. Minneapolis,
  Minnesota: Association for Computational Linguistics.

\bibitem[{Feng et~al.(2020)Feng, Guo, Tang, Duan, Feng, Gong, Shou, Qin, Liu,
  Jiang, and Zhou}]{feng2020codebert}
Feng, Z.; Guo, D.; Tang, D.; Duan, N.; Feng, X.; Gong, M.; Shou, L.; Qin, B.;
  Liu, T.; Jiang, D.; and Zhou, M. 2020.
\newblock CodeBERT: A Pre-Trained Model for Programming and Natural Languages.
\newblock arXiv:2002.08155.

\bibitem[{Fu et~al.(2017)Fu, Xu, Yu, and Yang}]{tf_fu2017wastk}
Fu, D.; Xu, Y.; Yu, H.; and Yang, B. 2017.
\newblock WASTK: a weighted abstract syntax tree kernel method for source code
  plagiarism detection.
\newblock \emph{Scientific Programming}, 2017.

\bibitem[{Goodfellow, Bengio, and Courville(2016)}]{goodfellow2016deep}
Goodfellow, I.; Bengio, Y.; and Courville, A. 2016.
\newblock \emph{Deep learning}.
\newblock MIT press.

\bibitem[{Guo et~al.(2021)Guo, Ren, Lu, Feng, Tang, Liu, Zhou, Duan,
  Svyatkovskiy, Fu, Tufano, Deng, Clement, Drain, Sundaresan, Yin, Jiang, and
  Zhou}]{guo2021graphcodebert}
Guo, D.; Ren, S.; Lu, S.; Feng, Z.; Tang, D.; Liu, S.; Zhou, L.; Duan, N.;
  Svyatkovskiy, A.; Fu, S.; Tufano, M.; Deng, S.~K.; Clement, C.; Drain, D.;
  Sundaresan, N.; Yin, J.; Jiang, D.; and Zhou, M. 2021.
\newblock GraphCodeBERT: Pre-training Code Representations with Data Flow.
\newblock arXiv:2009.08366.

\bibitem[{Hindle, Godfrey, and Holt(2009)}]{hindle2009s}
Hindle, A.; Godfrey, M.~W.; and Holt, R.~C. 2009.
\newblock What's hot and what's not: Windowed developer topic analysis.
\newblock In \emph{2009 IEEE international conference on software maintenance},
  339--348. IEEE.

\bibitem[{Husain et~al.(2020)Husain, Wu, Gazit, Allamanis, and
  Brockschmidt}]{husain2020codesearchnet}
Husain, H.; Wu, H.-H.; Gazit, T.; Allamanis, M.; and Brockschmidt, M. 2020.
\newblock CodeSearchNet Challenge: Evaluating the State of Semantic Code
  Search.
\newblock arXiv:1909.09436.

\bibitem[{Islam and Iqbal(2020)}]{tf_islam2020socer}
Islam, M.~M.; and Iqbal, R. 2020.
\newblock SoCeR: A New Source Code Recommendation Technique for Code Reuse.
\newblock In \emph{2020 IEEE 44th Annual Computers, Software, and Applications
  Conference (COMPSAC)}, 1552--1557. IEEE.

\bibitem[{Izadi, Heydarnoori, and Gousios(2021)}]{izadi2021topic}
Izadi, M.; Heydarnoori, A.; and Gousios, G. 2021.
\newblock Topic Recommendation for Software Repositories using Multi-label
  Classification Algorithms.
\newblock arXiv:2010.09116.

\bibitem[{Jones(1972)}]{jones1972statistical}
Jones, K.~S. 1972.
\newblock A statistical interpretation of term specificity and its application
  in retrieval.
\newblock \emph{Journal of documentation}.

\bibitem[{Kanade et~al.(2020)Kanade, Maniatis, Balakrishnan, and
  Shi}]{kanade2020learning}
Kanade, A.; Maniatis, P.; Balakrishnan, G.; and Shi, K. 2020.
\newblock Learning and Evaluating Contextual Embedding of Source Code.
\newblock arXiv:2001.00059.

\bibitem[{Karampatsis and Sutton(2020)}]{karampatsis2020scelmo}
Karampatsis, R.~M.; and Sutton, C. 2020.
\newblock SCELMo: Source Code Embeddings from Language Models.
\newblock arXiv:2004.13214.

\bibitem[{Kingma and Ba(2014)}]{kingma2014adam}
Kingma, D.~P.; and Ba, J. 2014.
\newblock Adam: A method for stochastic optimization.
\newblock \emph{arXiv preprint arXiv:1412.6980}.

\bibitem[{Landauer, Foltz, and Laham(1998)}]{landauer1998introduction}
Landauer, T.~K.; Foltz, P.~W.; and Laham, D. 1998.
\newblock An introduction to latent semantic analysis.
\newblock \emph{Discourse processes}, 25(2-3): 259--284.

\bibitem[{Le, Chen, and Babar(2020)}]{le2020deep}
Le, T.~H.; Chen, H.; and Babar, M.~A. 2020.
\newblock Deep learning for source code modeling and generation: Models,
  applications, and challenges.
\newblock \emph{ACM Computing Surveys (CSUR)}, 53(3): 1--38.

\bibitem[{Loshchilov and Hutter(2017)}]{loshchilov2017decoupled}
Loshchilov, I.; and Hutter, F. 2017.
\newblock Decoupled weight decay regularization.
\newblock \emph{arXiv preprint arXiv:1711.05101}.

\bibitem[{Navarro(2001)}]{navarro2001}
Navarro, G. 2001.
\newblock A Guided Tour to Approximate String Matching.
\newblock \emph{ACM Comput. Surv.}, 33(1): 31–88.

\bibitem[{Panichella et~al.(2013)Panichella, Dit, Oliveto, Di~Penta,
  Poshynanyk, and De~Lucia}]{2013genetics}
Panichella, A.; Dit, B.; Oliveto, R.; Di~Penta, M.; Poshynanyk, D.; and
  De~Lucia, A. 2013.
\newblock How to effectively use topic models for software engineering tasks?
  an approach based on genetic algorithms.
\newblock In \emph{2013 35th International Conference on Software Engineering
  (ICSE)}, 522--531. IEEE.

\bibitem[{Peters et~al.(2018)Peters, Neumann, Iyyer, Gardner, Clark, Lee, and
  Zettlemoyer}]{peters-etal-2018-deep}
Peters, M.~E.; Neumann, M.; Iyyer, M.; Gardner, M.; Clark, C.; Lee, K.; and
  Zettlemoyer, L. 2018.
\newblock Deep Contextualized Word Representations.
\newblock In \emph{Proceedings of the 2018 Conference of the North {A}merican
  Chapter of the Association for Computational Linguistics: Human Language
  Technologies, Volume 1 (Long Papers)}, 2227--2237. New Orleans, Louisiana:
  Association for Computational Linguistics.

\bibitem[{Puri et~al.(2021)Puri, Kung, Janssen, Zhang, Domeniconi, Zolotov,
  Dolby, Chen, Choudhury, Decker, Thost, Buratti, Pujar, and
  Finkler}]{puri21codenet}
Puri, R.; Kung, D.~S.; Janssen, G.; Zhang, W.; Domeniconi, G.; Zolotov, V.;
  Dolby, J.; Chen, J.; Choudhury, M.~R.; Decker, L.; Thost, V.; Buratti, L.;
  Pujar, S.; and Finkler, U. 2021.
\newblock Project CodeNet: {A} Large-Scale {AI} for Code Dataset for Learning a
  Diversity of Coding Tasks.
\newblock \emph{CoRR}, abs/2105.12655.

\bibitem[{Radford and Narasimhan(2018)}]{Radford2018ImprovingLU}
Radford, A.; and Narasimhan, K. 2018.
\newblock Improving Language Understanding by Generative Pre-Training.

\bibitem[{Raffel et~al.(2020)Raffel, Shazeer, Roberts, Lee, Narang, Matena,
  Zhou, Li, and Liu}]{raffel2020exploring}
Raffel, C.; Shazeer, N.; Roberts, A.; Lee, K.; Narang, S.; Matena, M.; Zhou,
  Y.; Li, W.; and Liu, P.~J. 2020.
\newblock Exploring the Limits of Transfer Learning with a Unified Text-to-Text
  Transformer.
\newblock arXiv:1910.10683.

\bibitem[{Rokon et~al.(2021)Rokon, Yan, Islam, and
  Faloutsos}]{rokon2021repo2vec}
Rokon, M. O.~F.; Yan, P.; Islam, R.; and Faloutsos, M. 2021.
\newblock Repo2vec: A comprehensive embedding approach for determining
  repository similarity.
\newblock In \emph{2021 IEEE International Conference on Software Maintenance
  and Evolution (ICSME)}, 355--365. IEEE.

\bibitem[{Salis et~al.(2021)Salis, Sotiropoulos, Louridas, Spinellis, and
  Mitropoulos}]{salis2021pycg}
Salis, V.; Sotiropoulos, T.; Louridas, P.; Spinellis, D.; and Mitropoulos, D.
  2021.
\newblock PyCG: Practical Call Graph Generation in Python.
\newblock arXiv:2103.00587.

\bibitem[{Sanh et~al.(2020)Sanh, Debut, Chaumond, and
  Wolf}]{sanh2020distilbert}
Sanh, V.; Debut, L.; Chaumond, J.; and Wolf, T. 2020.
\newblock DistilBERT, a distilled version of BERT: smaller, faster, cheaper and
  lighter.
\newblock arXiv:1910.01108.

\bibitem[{Sharma et~al.(2017)Sharma, Thung, Kochhar, Sulistya, and
  Lo}]{Sharma17}
Sharma, A.; Thung, F.; Kochhar, P.~S.; Sulistya, A.; and Lo, D. 2017.
\newblock Cataloging GitHub Repositories.
\newblock EASE'17, 314–319. New York, NY, USA: Association for Computing
  Machinery.
\newblock ISBN 9781450348041.

\bibitem[{Spinellis, Kotti, and Mockus(2020)}]{spinellis2020dataset}
Spinellis, D.; Kotti, Z.; and Mockus, A. 2020.
\newblock A dataset for github repository deduplication.
\newblock In \emph{Proceedings of the 17th international conference on mining
  software repositories}, 523--527.

\bibitem[{Svyatkovskiy et~al.(2020)Svyatkovskiy, Deng, Fu, and
  Sundaresan}]{Svyatkovskiy20}
Svyatkovskiy, A.; Deng, S.~K.; Fu, S.; and Sundaresan, N. 2020.
\newblock IntelliCode Compose: Code Generation Using Transformer.
\newblock In \emph{Proceedings of the 28th ACM Joint Meeting on European
  Software Engineering Conference and Symposium on the Foundations of Software
  Engineering}, ESEC/FSE 2020, 1433–1443. New York, NY, USA: Association for
  Computing Machinery.
\newblock ISBN 9781450370431.

\bibitem[{Theeten, Vandeputte, and Van~Cutsem(2019)}]{theeten2019import2vec}
Theeten, B.; Vandeputte, F.; and Van~Cutsem, T. 2019.
\newblock Import2vec: Learning embeddings for software libraries.
\newblock In \emph{2019 IEEE/ACM 16th International Conference on Mining
  Software Repositories (MSR)}, 18--28. IEEE.

\bibitem[{Thomas et~al.(2014)Thomas, Adams, Hassan, and
  Blostein}]{thomas2014studying}
Thomas, S.~W.; Adams, B.; Hassan, A.~E.; and Blostein, D. 2014.
\newblock Studying software evolution using topic models.
\newblock \emph{Science of Computer Programming}, 80: 457--479.

\bibitem[{Thung, Lo, and Jiang(2012)}]{thung2012detecting}
Thung, F.; Lo, D.; and Jiang, L. 2012.
\newblock Detecting similar applications with collaborative tagging.
\newblock In \emph{2012 28th IEEE International Conference on Software
  Maintenance (ICSM)}, 600--603. IEEE.

\bibitem[{Thung, Lo, and Lawall(2013)}]{thung2013automated}
Thung, F.; Lo, D.; and Lawall, J. 2013.
\newblock Automated library recommendation.
\newblock In \emph{2013 20th Working conference on reverse engineering (WCRE)},
  182--191. IEEE.

\bibitem[{Vaswani et~al.(2017)Vaswani, Shazeer, Parmar, Uszkoreit, Jones,
  Gomez, Kaiser, and Polosukhin}]{Vaswani17}
Vaswani, A.; Shazeer, N.; Parmar, N.; Uszkoreit, J.; Jones, L.; Gomez, A.~N.;
  Kaiser, L.; and Polosukhin, I. 2017.
\newblock Attention is All You Need.
\newblock In \emph{Proceedings of the 31st International Conference on Neural
  Information Processing Systems}, NIPS'17, 6000–6010. Red Hook, NY, USA:
  Curran Associates Inc.
\newblock ISBN 9781510860964.

\bibitem[{Washizaki et~al.(2019)Washizaki, Uchida, Khomh, and
  Gu{\'e}h{\'e}neuc}]{washizaki2019studying}
Washizaki, H.; Uchida, H.; Khomh, F.; and Gu{\'e}h{\'e}neuc, Y.-G. 2019.
\newblock Studying software engineering patterns for designing machine learning
  systems.
\newblock In \emph{2019 10th International Workshop on Empirical Software
  Engineering in Practice (IWESEP)}, 49--495. IEEE.

\bibitem[{Yi and Allan(2009)}]{yi2009comparative}
Yi, X.; and Allan, J. 2009.
\newblock A comparative study of utilizing topic models for information
  retrieval.
\newblock In \emph{European conference on information retrieval}, 29--41.
  Springer.

\bibitem[{Yujian and Bo(2007)}]{YujianLevenshtein}
Yujian, L.; and Bo, L. 2007.
\newblock A Normalized Levenshtein Distance Metric.
\newblock \emph{IEEE Transactions on Pattern Analysis and Machine
  Intelligence}, 29(6): 1091--1095.

\bibitem[{Zhang et~al.(2017)Zhang, Lo, Kochhar, Xia, Li, and
  Sun}]{zhang2017detecting}
Zhang, Y.; Lo, D.; Kochhar, P.~S.; Xia, X.; Li, Q.; and Sun, J. 2017.
\newblock Detecting similar repositories on GitHub.
\newblock In \emph{2017 IEEE 24th International Conference on Software
  Analysis, Evolution and Reengineering (SANER)}, 13--23. IEEE.

\bibitem[{Zhang et~al.(2019)Zhang, Xu, Li, Meng, Wang, Li, and
  Han}]{zhang2019higitclass}
Zhang, Y.; Xu, F.~F.; Li, S.; Meng, Y.; Wang, X.; Li, Q.; and Han, J. 2019.
\newblock Higitclass: Keyword-driven hierarchical classification of github
  repositories.
\newblock In \emph{2019 IEEE International Conference on Data Mining (ICDM)},
  876--885. IEEE.

\end{thebibliography}
\newpage
\appendix
\section{Details of the Github Crawler}
GitHub repositories are often classified by its owner using user-defined topics, which can contain abbreviations, typos, and repetitions. Because of the large variations in topic names, GitHub also defines 480 featured topics, a limited number of predefined topics to be associated with the repository by its owner. In order to have a standard label set, the crawler maps the user-defined topics by user to the GitHub featured topics using string matching method \cite{cohen2011fuzzywuzzy, navarro2001} relying on  threshold (90\%) of "Fuzzy matching" (Levenshtein ratio \cite{YujianLevenshtein}) distance of words to identify similar topics.   Figure~\ref{fig:featured_topics}a in the paper compares the distributions of the number of topics associated to a repository before and after the mapping to featured topics. The decrease in the number of repositories with a given number of tags in  Fig.~\ref{fig:featured_topics}a shows that GitHub users have a tendency to associate numerous topics, which can actually be very similar, to their repository in order to increase their visibility in the GitHub search engine. Note that most of the user-defined topics are effectively related to a featured topic. Furthermore, a significant amount (approximately 32\%) of repositories have been found to not be associated with any of the topics and this number increases when restricting to featured-topics. Topical can thus assist the user to automate the process of generating topic tags amongst a limited set. 
\section{Dependency Graph Embedding}
\begin{figure}[t!]
\centering
\subfloat[]{\includegraphics[scale=0.5]{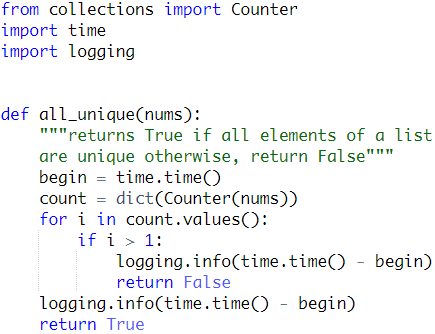}}\quad
\subfloat[]{\includegraphics[scale=0.5]{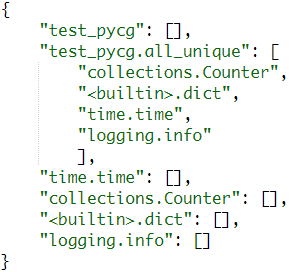}}\quad
\subfloat[]{
\begin{minipage}{.4\textwidth}
\centering
\includegraphics[scale=0.6]{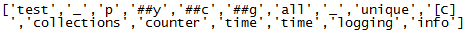}\\
\includegraphics[scale=0.6]{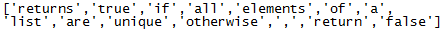}\\
\includegraphics[scale=0.6]{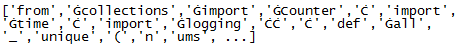}
\end{minipage}}
\caption{(a) Code snippet. (b) PyCG output dependencies graph on the code snippet. (c) Tokenized sequences for PyCG dependencies, docstring and source code.}
\label{fig:tokenized}
\end{figure}

In order to generate a dependency graph embedding, we follow the steps below: 
\begin{itemize}
    \item Assume we have a script with methods $M = \{m_1, m_2, ..., m_n\} $ and their respective imported uses $I = \{i_1, i_2, ..., i_n\} $. Using PyCG, for each script, we retrieve its corresponding edges that link all methods in a script to other classes and other methods, implemented in the same repository but also in external library calls as presented in Fig.~\ref{fig:tokenized}a-b.
    \item In order to tokenize a dependency graph into the DistilBERT model, we dedicate a special token to indicate the link between two nodes in the graph, a method name and its class imports or other method usage. That is: 
    \begin{itemize}
        \item Using import dependencies, we tokenize input $X = \{[CLS], M, [SEP], I\}$  where $[CLS]$ is a special token for classification and $[SEP]$ is the special token to delimit both information.
        \item Using the rest of the dependency graph, we tokenize input 
        \begin{equation}
            X = \{[CLS], m_1, [C], i_1, [SEP], m_2, [C], i_2, [SEP],..\}
        \end{equation}where $[CLS]$ is a special token for classification, $[C]$ is our introduced special token indicating an edge and $[SEP]$ is the special token to delimit the pairs.
    \end{itemize}
    \item All first rank edges of the graph are then concatenated sequentially with the separating token, before being passed to the DistilBERT model. The usage of the path as a sentence in a trained natural language model is assumed to be desirable as we expect to obtain embedding with "some" relation to the distance between function calls (also words).
\end{itemize}
Figure~\ref{fig:tokenized}b presents how we retrieve the nodes from the PyCG output and Fig.~\ref{fig:tokenized}c presents how we tokenize them after introducing $[C]$ as a DistilBERT special token. 



\section{Docstring Embedding}
Although GraphCodeBERT also processes comments in code, here we designate a separate embedding for retrieving textual information from comments and pre-processed method names. This allows us to pre-process textual information found in source code to target repository-level tasks. We extract docstrings, function textual names and file names. 
In order to obtain a fixed-size final embedding, we integrate file names and method names into a single sentence and separate them from the script docstrings using the special token for separation and encode the tokenized input in DistilBERT \cite{sanh2020distilbert} (a natural language embedding, which is a pre-trained model on English language). Similar to code embedding, we concatenate comments and function names from a script and use maximum of 512 tokens size as vector input for DistilBERT.

\section{Additional Results}
\subsection{Repository Topics}
\begin{table}[t]
\centering
  \caption{Topics and size of the dataset (number of repository) for the reported experiments. A 70-30 
  \% train-test split is used to generate training and test data. The metrics reported/shown in the figures are computed on the test data.}
  \label{tab:topics}
  \begin{tabular}{p{0.2\linewidth} p{0.2\linewidth} p{0.5\linewidth}}
    \hline
    No. of topics & Number of Repositories & Selected topics\\
    \hline
    5 & 760 &  Machine Learning (ML), Deep Learning (DL), Database, Django, Reinforcement Learning (RL)\\
    10 & 1200 & ML, DL, Database, Django, RL, Tensorflow(TF), Ethereum, Computer Vision (CV), Bot, Hacktoberfest\\
    15 & 1376 & ML, DL, Database, Django, RL, TF, Ethereum, CV, Bot, Hacktoberfest, Natural Language Processing (NLP), Algorithm, Bitcoin, Cryptocurrency, Flask\\
    20 & 1586 & ML, DL, Database, Django, RL, TF, Ethereum, CV, Bot, Hacktoberfest, NLP, Algorithm, Bitcoin, Cryptocurrency, Flask, Security, Docker, Linux, API, Covid-19 \\
  \hline
\end{tabular}
\end{table}

\subsection{Low Dimensional Visualizations}
\begin{figure}[t!]

\subfloat[]{\includegraphics[scale=0.55]{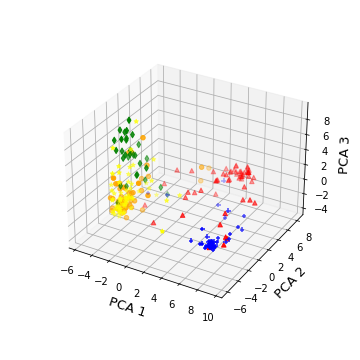}}\quad
\subfloat[]{\includegraphics[scale=0.5]{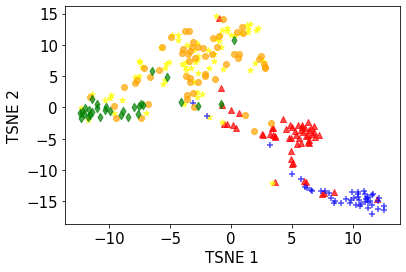}\llap{\shortstack{%
        \includegraphics[scale=.27]{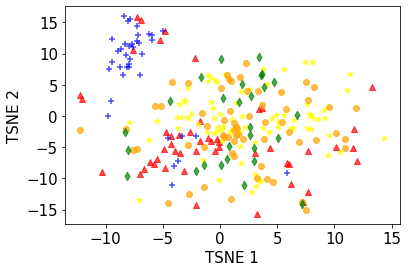}\\
        \rule{0ex}{1.3in}%
     }
 \rule{-0.2in}{0ex}}
}
\caption{(a) 3D PCA projection of repositories embedding for the top 5 topics by frequency. (b) 2D TSNE projection of repositories embedding for the top 5 topics. Inset shows the 2D TSNE projection for embedding for the top 5 topics without the attention mechanism. The top 5 topics: ML, DL, Database, Django, and RL are represented by Orange, Yellow, Red, Blue and Green points respectively.}
\label{embedding}
\end{figure}
Figure \ref{embedding} introduces a latent visualization of the repository embedding from Topical, using the attention model.  We embed repositories from 5 popular topics (see Table \ref{tab:topics}) crawled directly from GitHub. We use TSNE projection on a 2D latent space, and the top three PCA components for 3D projection. Figure \ref{embedding} shows a clear separation between embedding projections from Topical in the task for classification of GitHub topics. 
It is worth noting that two latent components do not fully capture the abstract differences between repositories, such as the context or topic of the repository. However separation is still clearly more pronounced (Figure \ref{embedding}b) with the Topical embedding in comparison to an embedding achieved without the attention mechanism (the inset of Figure \ref{embedding}b).

The PCA projection (Fig.~\ref{embedding}a) also validates, at least visually, our decision to reduce dimensionality of the embedding, before the classifier head. The inset of Fig.~\ref{embedding}b shows also the projection of the embedding without the attention mechanism. It shows that using the mean embedding (\ie~no attention mechanism) can be useful but far less efficient in clustering similar embedding of repositories in comparison to the proposed attention model.

\section{Ablation Studies}
\subsection{Effect of changing embedding components}
In our second ablation study, we investigate the individual importance of the script, dependency, and docstring embeddings towards the topic classification task. In particular, we remove each component one at a time and test the resulting performance of the downstream topic classifier - absence of the most important component should cause the biggest drop in performance. From, Table~\ref{tab:component_ablation}, we see that the code embedding is likely the most important, with a $F_{1}$ performance drop of $\approx$ 0.035 upon removal. Surprisingly, the dependency graph embedding turns out to be less important for the topic classification task, with its removal providing no losses (rather, statistically insignificant gains) in performance. We hypothesize that this is likely because information pertaining to topic classification is contained in code and docstrings, with likely redundancies in the dependency graph. We leave investigation of ablative performance on other downstream tasks (e.g. information retrieval, where the dependency embeddings may play a more significant role) to future work.
\begin{table}[H]
\centering
\resizebox{\columnwidth}{!}{%
\begin{tabular}{ccc}
\hline
\textbf{Removed Component} & \textbf{$F_{1}$}       & \textbf{LRAP}     \\ \hline
None                       & 0.661 $\pm$ 0.015 & 0.791 $\pm$ 0.003 \\
Code Embedding             & 0.626 $\pm$ 0.004 & 0.781 $\pm$ 0.009 \\
Docstring Embedding        & 0.639 $\pm$ 0.010 & 0.790 $\pm$ 0.005 \\
Dependency Graph Embedding & \textbf{0.665 $\pm$ 0.014} & \textbf{0.793 $\pm$ 0.005} \\ \hline
\end{tabular}%
}
\caption{Ablation study on different components of the repository embedding}
\label{tab:component_ablation}
\end{table}
\subsection{Effect of changing number of sampled scripts}
\begin{figure}[H]
    \centering
    \includegraphics[scale=0.5]{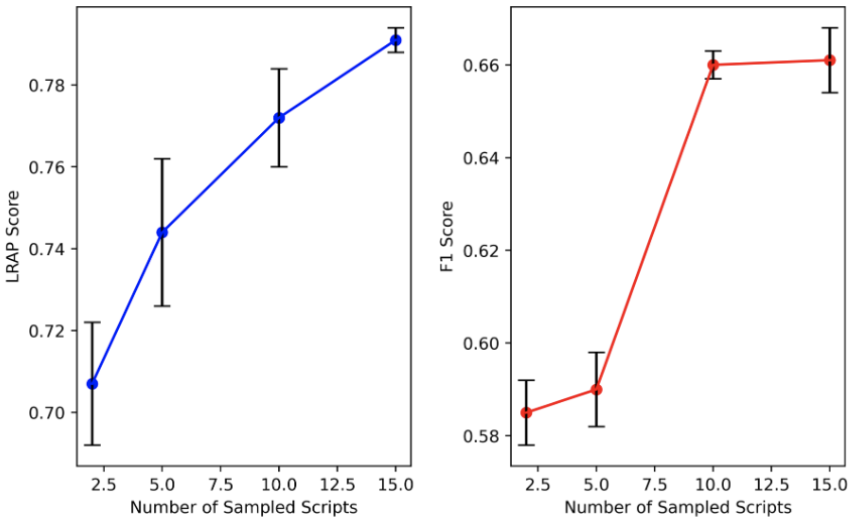}
    \caption{Number of scripts sampled for the code embedding vs downstream classifier performance}
    \label{fig:sampled_scripts}
\end{figure}

From the previous section, we saw the importance of the script embedding in generating the final repository representation. In this section, we ask the question - how sensitive is final performance of the classifier to the number of sampled scripts used to create the script embedding. We investigate different numbers of sampled scripts $\in \{2,5,10,15\}$ in Figure \ref{fig:sampled_scripts}. The performance is clearly shown to increase with the number of sampled scripts, however, there is a clear plateau. We also found that it becomes increasingly computationally expensive to compute embeddings with a large number of sampled scripts - there is thus a performance-cost tradeoff. 
\subsection{Effect of changing architectural parameters}
In this section, we investigate the two components of the embedding generation procedure: 
\begin{itemize}
    \item Compression of the higher dimensional code, docstring, and dependency embedding into a lower dimensional space. For this, we compare PCA with a simple linear layer with weight matrix of shape $768 \times 64$ (where $768$ is the size of each high dimensional embedding and $64$ is the size of the corresponding dimensionally reduced embedding). 
    \item Encoding of the concatenated lower dimensional embeddings and subsequent application of the attention mechanism (i.e. part $b$ in Figure \ref{fig:model_scheme}). We replace the RNN sequence encoder with alternatives in Table \ref{tab:model_ablation} and measure the resulting impact on classification performance ($20$ topics). 
\end{itemize}
\begin{table}[H]
\centering
\resizebox{0.8\columnwidth}{!}{%
\begin{tabular}{ccc}
\hline
\textbf{Encoder Model} & \textbf{$F_{1}$}                    & \textbf{LRAP}                  \\ \hline
Bi-GRU                 & 0.661 $\pm$ 0.015 & \textbf{0.791 $\pm$ 0.003} \\
Bi-LSTM                & \textbf{0.664 $\pm$ 0.018} & 0.786 $\pm$ 0.016 \\
MLP            & 0.629 $\pm$ 0.009 & 0.762 $\pm$ 0.010 \\
\hline
\end{tabular}%
}
\caption{Ablation study of the Encoder in Figure \ref{fig:model_scheme} b). We used a standard Multilayer Perceptron (MLP) that performs the same job as the GRU sequence encoder, i.e. mapping a tensor of shape $(*,*,192)$ to a shape $(*,*,96)$. We also chose the architecture to have approximately the same number of parameters as the GRU to keep the study fair.}
\label{tab:model_ablation}
\end{table}
\begin{table}[H]
\centering
\resizebox{0.8\columnwidth}{!}{%
\begin{tabular}{ccc}
\hline
\textbf{Dim Reduction} & \textbf{$F_{1}$}     & \textbf{LRAP}   \\ \hline
PCA                           & \textbf{0.661 $\pm$ 0.015} & \textbf{0.791 $\pm$ 0.003} \\
Linear Layer                  & 0.624 $\pm$ 0.018 & 0.788 $\pm$ 0.003 \\
\hline
\end{tabular}%
}
\caption{Ablation study on embedding compatification techniques}
\label{tab:embedding}
\end{table}
Our observations are the following
\begin{itemize}
    \item The hidden states of bidirectional sequence encoder models provide a better embedding representation than that of an MLP. 
    \item Using PCA for dimensionality reduction provides better results than having a learned linear layer that outputs a representation of reduced dimension. This may be because PCA already yields components that explain the variance in the data the most.
\end{itemize}



\end{document}